\documentclass{article}
\pdfoutput=1
\usepackage{amssymb,amsmath}
\usepackage{xcolor}
\usepackage[dvips]{graphicx}
\usepackage{float}
\usepackage{cite}
\usepackage{caption}
\usepackage{typearea}
\typearea{12}

\begin{document}

\Large{\begin{center} Annihilation of Secluded Dark Matter into $W^+W^-$ \\ Enhanced by $P$-wave Sommerfeld Effect \end{center}}
{\normalsize \begin{center}Nobuki Yoshimatsu \end{center}}
{\it \footnotesize \begin{center}Supreme School for Advanced Education\\1-5-4\ Nishi Takamatsu, Wakayama-shi, Wakayama 641-0051, Japan\\ Chiben Gakuen Wakayama Junior/Senior High School \\ 2066-1\ Fuyuno, Wakayama-shi, Wakayama 640-0392, Japan\\{\footnotesize E-mail address:\ nyoshimatsu260@gmail.com}\end{center}}\small{\begin{center}Abstract \end{center} We propose that a pair of annihilation of the secluded dark matter may accommodate a source of the halo gamma ray signal reported recently, through the $p$-wave Sommerfeld enhancement. We show that given a weakly coupling of the dark sector to the Higgs bosons, the dark matter annihilation into $W^+W^-$, even though induced at 1-loop level, is desirably amplified. We also argue that the supersymmetric framework may readily embody such a model within rather minimal contents. {\flushleft   }}

{\begin{center} {\bf Introduction} \end{center}}

{\flushleft \ }
Dark matter (DM) theory opened up a variety of prospects for cosmology and astronomy \cite{Zwicky,Cirelli:2024ssz}. One remarkable notion, as argued by \cite{Peebles}, is that the DM plays essential roles in forming the large-scale structure of the universe. Furthermore, the DM relics seemingly comprise about one-fourth of the mass density of the universe, which is over five times larger than that of the ordinal particles. Based upon particle physics, particularly quantum field theory, the weakly interacting massive particle (WIMP) has been recognized as a promising candidate for the DM \cite{Arbey,Cirelli:2024ssz,Arcadi:2017kky,Roszkowski}, although there is no direct evidence for its appearance at either colliders or underground experiments. 
Recently, the author of \cite{Totani:2025fxx} studied 15 years of the Fermi LAT data to search for annihilation gamma rays from dark matter in the Milky Way halo, and found a spectral peak
around 20 GeV of the photon energy. It was then reported that the excess implies that the WIMPs, with the mass of $500-800$ GeV, annihilate into $W^{+}W^-$ or $b \bar{b}$, where the cross section is $\left<\sigma v_{rel} \right>_{MW} \sim (5-8) \times 10^{-25} \ cm^3/s$. In line with it, the author of \cite{Totani:2025fxx} pointed out that the cross section might be incompatible with the upper limitations derived from dwarf spheroidal galaxies \cite{Fermi-LAT:2015att, Fermi-LAT:2016uux, MAGIC:2020ceg, McDaniel:2023bju}, and is more than an order of magnitude larger compared to the canonical thermal relic cross section (postulated the $s$-wave annihilation of the DM). Subsequently, it was suggested in \cite{Murayama:2025ihg} that resonant dark matter may give a reasonable explanation to these concerns.

In this letter, we aim to address the dilemma by proposing an alternative within the context of the Sommerfeld enhancement through the $p$-wave annihilation, which exhibits much more sensitive dependence on the DM velocity than the $s$-wave \cite{Beneke:2024iev, Chen:2024njd, Kaplinghat:2013yxa, Kainulainen:2015sva, Kahlhoefer:2017umn, Hufnagel:2018bjp, Hambye:2019tjt, Biondini:2021ccr,Biondini:2021ycj } (See \cite{Lattanzi:2008qa, Hisano:2003ec,March-Russell:2008klu} for the $s$-wave Sommerfeld enhancement). Specifically, we investigate whether the "secluded dark matter model" \cite{Pospelov:2007mp} may provide the DM annihilation favored by observations on both Milky Way halo and dwarf galaxies. Regarding, it has been pointed out that the sort of model readily accounts for the observed DM abundance, using a Yukawa coupling in the dark sector, whereas the DM itself is unlikely to be detected either directly or indirectly. As a minimal extension, we suppose a weakly coupling of the boson to the Higgs doublet, examining the consequences brought by the Sommerfeld enhancement. First, we demonstrate that the stable fermion due to a $Z_2$ symmetry has the correct thermal relics for its mass of $\sim 800$ GeV. Further, we find that a pair of DM annihilation into two Higgs bosons, which is induced by a 1-loop effect, can be highly enhanced at $v=100-200$ km/s through the $p$-wave Sommerfeld effect, so that the cross section reported by \cite{Totani:2025fxx} is embodied at Milky Way halo. In contrast, it is verified that the annihilation cross section at $v\sim 10$ km/s is even smaller than upper limitations on dwarf galaxies. Finally, we roughly illustrate that the supersymmetric framework may support the viability of such a model, providing the coupling structure and the mass of the dark sector particles.

{\begin{center} {\bf Explicit model} \end{center}}

\noindent
Let us introduce one Majorana fermion (denoted $\chi$) and one real scalar (denoted $\phi$), both of which are gauge singlets. We then consider the following Lagrangian density:
\begin{align}
\mathcal{L} \supset \ & i \chi \gamma^{\mu }\partial_{\mu}\chi+ \dfrac{1}{2}\partial^{\mu}\phi\partial_{\mu}\phi   -m_{\chi}\chi \chi -\dfrac{1}{2} m^2_{B} \phi^2 \notag \\ & +y \phi \chi \chi+ \dfrac{\lambda}{2} \phi^2 \left|H\right|^2+M\phi\left|H\right|^2, \label{1}
\end{align}
where $H^t=(H^+, H^0)$ denotes the SM Higgs doublet. $y, \lambda$ are dimensionless coupling constants which are assumed to be real-valued. $m_{\chi}$ is the mass of $\chi$, while $m_B, M$ denote mass parameters.
Here we assume a matter parity, i.e. a $Z_2$ symmetry. 
Each charge assignment is shown in Table 1.
\begin{table}
\centering
    \caption{Charge assignment}
    \label{tab:hogehoge}
\begin{tabular}{c|c|c|c}
 \hline 
& $\chi$  & $\phi$ & $H$ \\ \hline
 $Z_2$  & $-$ & $+$ & $+$ \\ \hline
\end{tabular}
\end{table}
Note that $\chi$ has the odd parity and thus can be stable, which is identified with a DM candidate. Also, we omit $\phi^4, \left|H\right|^4$ or higher-order terms which are irrelevant to our analyses.
Besides, we assume that $M$ is small enough to suppress the $\phi-H^0$ mixing. In this model, it is understood that a Yukawa coupling in Eq.(\ref{1}) entails the Sommerfeld effect at a pair of $\chi$ annihilation. Particularly, we focus upon the $p$-wave enhancement that occurs in the processes of our interest, i.e., $\chi \chi \rightarrow \phi \phi$ or $\chi \chi \rightarrow W^+ W^{-}$. 

{$\bullet$ {\it In freeze-out}}

\noindent
We evaluate the relic abundance of $\chi$ at the freeze-out temperature. A pair of $\chi$ annihilates into $\phi \phi$ through $\chi$ exchange. Notice that the $p$-wave yields the leading contribution to the cross section, which is given by
\begin{align}
 \sigma^{ann}_0 v \simeq \dfrac{3 \pi \alpha^2_{y} \left|\vec{p}\right|^2}{8 m^2_{\chi}},
\end{align}
at the tree level. Here, $\alpha_y = y^2/4 \pi$ and $\vec{p}$ is $3$-momentum of the center-mass-frame of an incoming $\chi$ (and hence the other $\chi$ carries $-\vec{p}$). %and $E^2=m^2_{\chi}+\left|\vec{p}\right|^2$.
We then obtain the thermally averaged cross section: 
\begin{align}
 \left<\sigma^{ann} v_{rel}\right>_0 & \simeq \dfrac{1}{\left(\sqrt{2 \pi} v_0\right)^3} \int d^3v \sigma_{ann} v e^{-v^2/2 v^2_0} \notag \\ &= \dfrac{3 \pi \alpha^2_{y}}{8 m^2_{\chi}} \cdot \dfrac{6}{x},
\end{align}
where $x\equiv m_{\chi}/T=3 v_0^{-2}$. At this point, noting that $\phi$ is in thermal equilibrium around (or below) the reheating temperature through the interactions with the Higgs bosons in Eq.(\ref{1}) \cite{1}, we observe that $\chi$ can be thermalized as well. Taking account of the Sommerfeld factor (denoted $S_{l=1}\left(v\right)$), we write the cross section as follows:
\begin{equation}
    \left<\sigma^{ann} v_{rel}\right> = S_{l=1}\left(v \right)\left<\sigma^{ann} v_{rel}\right>_0.
\end{equation} 
At the freeze-out temperature, $x$ is expressed as 
\begin{align}
    x_f & \simeq \ln{\left[\dfrac{9 \pi \alpha^2_{y}g}{4 \cdot \left(2 \pi\right)^{3/2}} \left(\dfrac{90}{\pi^2 g_{\ast}}\right)^{1/2} \dfrac{M_{pl}}{m_{\chi}}\right]} \notag \\ &-\dfrac{1}{2} \ln{\left(\ln{\left[\dfrac{9 \pi \alpha^2_{y}g}{4 \cdot \left(2 \pi\right)^{3/2}} \left(\dfrac{90}{\pi^2 g_{\ast}}\right)^{1/2} \dfrac{M_{pl}}{m_{\chi}}\right]}\right)}. 
\end{align}
Here, we set $S_{l=1}\left(v = (3/x)^{1/2}\right) = 1$ for simplicity because the Sommerfeld effect is considered negligible around $x_f$. One then derives the thermal relics of $\chi$:
\begin{align}
  & \dfrac{\rho_{\chi}}{s} \simeq 0.145 \dfrac{g}{g_{\ast s}} m_{\chi} x_f^{3/2} e^{-x_{f}} \notag \\ &\ \ = 0.145 \dfrac{g}{g_{\ast s}} m_{\chi} x^2_f \left[\dfrac{9 \pi \alpha^2_{y}g}{4 \cdot \left(2 \pi\right)^{3/2}} \left(\dfrac{90}{\pi^2 g_{\ast}}\right)^{1/2} \dfrac{M_{pl}}{m_{\chi}}\right]^{-1}.
\end{align}
Eventually, it follows that $\rho_{\chi}/s \simeq 4.6 \times 10^{-10}$ GeV and $x_f \simeq 27.4$ for $m_{\chi}=800$ GeV and $\alpha_y=0.07$, for which the DM density is found to dominantly comprise $\chi$. Here, we set $g_{\ast}=g_{\ast s}=90$ at $x_f$ \cite{Husdal:2016haj}.

{$\bullet$ {\it Annihilation into $W^+ W^{-}$}}

\noindent

We examine annihilation of $\chi \chi \rightarrow H^+ H^{+ \ast}$, converted to $\chi \chi \rightarrow W^+ W^{-}$ in the limit of $m_{\chi} \gg m_{W}$. The cross section is as follows:
\begin{equation}
    \sigma^{ann} v_{rel} = S_{l=1}\left(v \right)\sigma^{ann}_0 v_{rel}.
\end{equation} 
where $\sigma_0 v_{rel}$, that is the cross section without the Sommerfeld effect, is given by 
\begin{align}
    \sigma_0 v_{rel} \simeq \dfrac{1}{8 \pi m^2_{\chi}} \cdot \dfrac{\alpha_y^2 \lambda^2}{16 \pi^2} \cdot \dfrac{\left|\vec{p}\right|^2}{m^2_{\chi}} \cdot f\left(\dfrac{m^2_{\phi}}{m^2_{\chi}} \right),
\end{align}
and $f\left( t\right)$ is expressed as
\begin{align}
    f\left( t\right) =  \int_0^1 dx \int_0^{1-x} dy  \times \dfrac{\left(2-y \right)^2}{\left[\left(1-2x-y\right)^2+2y-1-4x(1-x-y)v^2+\left(1-y\right)t \right]^2}.
\end{align}
Let us then estimate the magnitude of $\sigma^{ann} v_{rel}$. According to \cite{Beneke:2024iev}, close to the first quasi-bound state that appears at $\epsilon_{\phi} \equiv m_{\phi}/(\alpha_y m_{\chi})\simeq 0.11$, the Sommerfeld factor is numerically estimated as
\begin{equation}
S_{l=1} = \dfrac{B_{21}}{\left(\epsilon^2_v-\epsilon^2_{21}\right)^2+ C^2_{21} \epsilon^6_{21}/4},
\end{equation}
where $\epsilon_v =v/(2\alpha_y), B_{21}=0.0209, C_{21}=26.4$ and $\epsilon_{21} \simeq 3.5 \times 10^{-3}$ is the dimensionless energy of the first quasi-bound state \cite{qbs}.
Postulated $\lambda \simeq 10^{-3}$,
%$m \simeq 5.0$ GeV
  $m_{B} \simeq 2.77$ GeV and $m_{\chi}=800$ GeV as instant values, the effective mass of $\phi$ is given by 
 \begin{equation}
    m_{\phi}=\sqrt{m^2_{B}+\lambda \left<H^0 \right>^2} \simeq 6.16 \ \text{GeV},
 \end{equation}
 realizing the first quasi-bound state, i.e. $\epsilon_{\phi}\simeq 0.11$ for $\alpha_y =0.07$. Here $\left<H^0 \right> =174$ GeV is the vacuum expectation value (VEV) developed by $H^0$ due to the electroweak symmetry breaking (EWSB). As displayed in Figure 1, one then obtains the annihilation cross section of $\chi \chi \rightarrow W^+ W^{-}$:
 \begin{equation}
  \sigma^{ann} v_{rel} \simeq 7.1 \times 10^{-25} cm^3/s \label{cross-sec1}
 \end{equation}
 at $\epsilon_{v} \simeq 3.5 \times 10^{-3}$ which corresponds to the DM velocity of $\sim 147$ km/s. (At this point, we note that $f(m^2_{\phi}/m^2_{\chi})$ takes an approximately constant value of $\sim 2.0 \times 10^6$ for $v \lesssim 10^{-3}$.) We thus conclude that the $p$-wave enhancement well explains the cross section at Milky Way halo reported by \cite{Totani:2025fxx}.  
 \begin{figure}
    \caption{The DM annihilation cross section, including the $p$-wave Sommerfeld effect, is displayed. \ \ \ \ \ \ Here, we set $f(m^2_{\phi}/m^2_{\chi})=2.0 \times 10^6$.} 
   \begin{flushright}
\includegraphics[width=\textwidth]{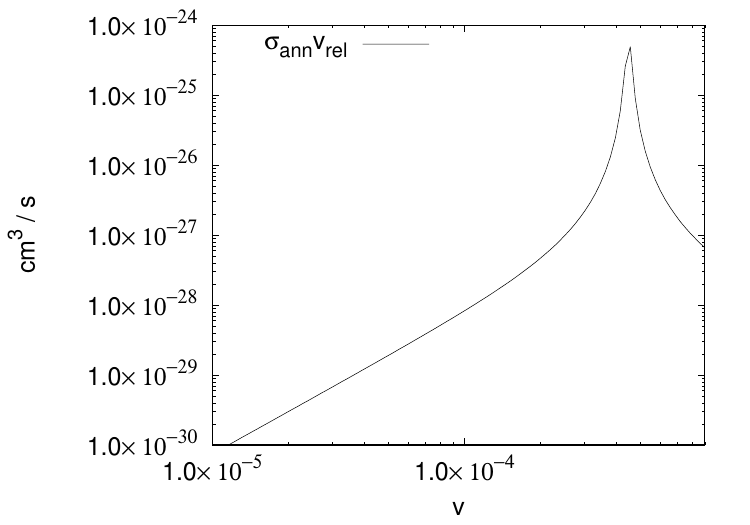}
    \end{flushright}
    \caption*{   \ \ \ \ \ }
\end{figure}
 %$f(m^2_{\phi}/m^2_{\chi})\simeq 1.3 \times 10^4$.  
 In contrast, the DM annihilation is highly suppressed at the DM velocity of $\mathcal{O}(10)$ km/s:
 \begin{equation}
    \sigma^{ann} v_{rel} \simeq 4.1 \times 10^{-29} \left(\dfrac{v}{10^{-4}}\right)^2 \ cm^3/s, \label{cross-sec2}
 \end{equation}
which is far less than the upper limitations on dwarf galaxies \cite{Fermi-LAT:2015att, Fermi-LAT:2016uux, MAGIC:2020ceg, McDaniel:2023bju}. Eqs.(\ref{cross-sec1}) and (\ref{cross-sec2}) are also expected to indicate the cross section of $\chi \chi \rightarrow H^0 H^0, ZZ$.

 {$\bullet$ \it $\sigma v (\chi \chi \rightarrow \phi \phi)$ at $p$-wave enhancement}
 \begin{figure}
    \caption{$\Gamma_{ann}/H$ is peaked at $\epsilon_v \simeq \epsilon_{21}$, although the DM remains out of thermal equilibrium.} 
   \begin{flushright}
\includegraphics[width=\textwidth]{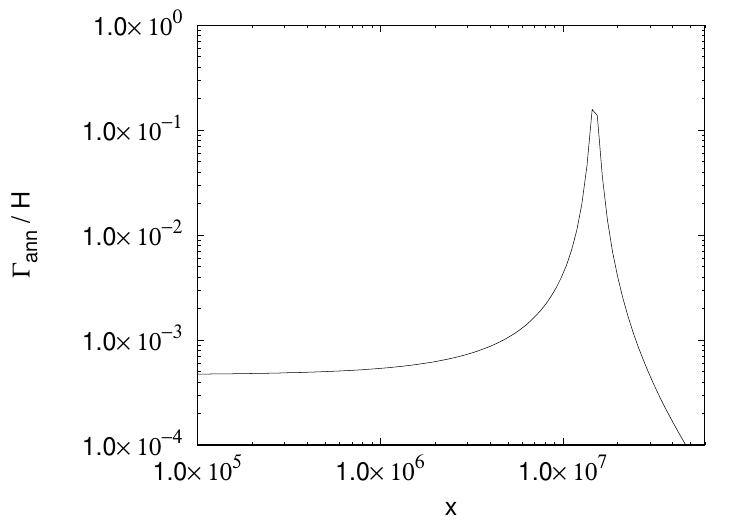}
    \end{flushright}
    \caption*{   \ \ \ \ \ }
\end{figure}
 \noindent
 
 Subsequently, we investigate the cross section of $\chi \chi \rightarrow \phi \phi$ enhanced by the $p$-wave Sommerfeld effect. After the freeze-out, the ratio of the annihilation rate to the Hubble parameter is expressed as
 \begin{equation}
    \dfrac{\Gamma_{ann}}{H} = S_{l=1} \cdot \left(\dfrac{x_f}{x}\right)^2.
 \end{equation} 
Hence, it is found that $\Gamma_{ann}/H \simeq 0.22$ around the peak of $S_{l=1}$.
We therefore observe that $\chi$ remains out of thermal equilibrium below the freeze-out temperature (See Figure 2). 
{\flushleft \ }

\begin{center}{\bf Decay of $\phi$} \end{center}

\noindent
We discuss the fate of $\phi$. As seen from Eq.(\ref{1}), there exists a mixing term with $H^0$:
\begin{equation}
    V \supset M \left<H^0 \right> \phi H^0 +h.c.,
\end{equation}
leading to the mixing angle of $\epsilon \simeq M\left<H^0 \right>/m^2_{H^0}$. If $\epsilon$ is taken as the allowed level by the LHC experiments \cite{lhc-higgs}, one obtains the decay rate of $\phi$ into $\tau^+ \tau^-$:
\begin{align}
    \Gamma_{\phi \rightarrow \tau^+ \tau^-} & \simeq \dfrac{1}{8 \pi} \epsilon^2 \left(\dfrac{m_{\tau}}{\left<H^0 \right>}\right)^2 \cdot m_{\phi} \cdot \left(1-\dfrac{4m_{\tau}^2}{m_{\phi}^2}\right)^{3/2} \notag \\ &= 2.7 \times 10^{-19} \left(\dfrac{\epsilon}{10^{-7}}\right)^2 \ \text{GeV}.
\end{align}
that corresponds to the lifetime of
\begin{equation}
    \tau_{\phi} \simeq 2.4 \times 10^{-6} \left(\dfrac{\epsilon}{10^{-7}}\right)^{-2} s.
\end{equation}
It is thus verified that $\phi$ decays well before Big Bang Nucleosynthesis (BBN).

\begin{center}{\bf Discussion}\end{center}

\noindent
Here, we roughly discuss a possibility that the sort of model may be made viable in the framework of supersymmetry (SUSY). Let us suppose the following superpotential and K$\ddot{a}$hler potential:
\begin{align}
    & W = \dfrac{a}{3!} \Phi^3+b \Phi H_uH_d+ c \dfrac{S}{M_{pl}} \Phi H_uH_d, \notag \\ & K= \dfrac{\kappa_1 S^{\dagger}}{2 M_{pl}} \Phi^2+ \dfrac{\kappa_2 \left|S\right|^2}{M_{pl}^2} \left|\Phi \right|^2+ \dfrac{\kappa_3 \left|S\right|^2}{2 M_{pl}^2} \Phi^2+h.c..
\end{align}
Among the Lagrangian density are then the relevant terms:
  \begin{align}
\mathcal{L} \supset &\ a \eta \bar{\psi} \psi + \left|b\right|^2 \left|\eta \right|^2 \left(\left|H_u\right|^2+\left|H_d\right|^2\right) \notag \\ & -\sqrt{3} \kappa_1 m_{3/2} \bar{\psi} \psi - 3Re\left(\kappa_2\right) m_{3/2}^2 \left|\eta \right|^2 \notag \\ & -\dfrac{3\kappa_3}{2} m_{3/2}^2 \eta^2  +\sqrt{3} c m_{3/2} \eta H_uH_d \notag \\ &+\dfrac{ab}{2} \eta^2 H_u^{\ast} H_d^{\ast} + h.c..
\end{align}
where $a, b, c$ and $\kappa_i \ (i=1-3)$ denote the dimensionless coupling constants. $\Phi=\eta+ \theta \psi$ is a gauge singlet chiral superfield, and thus $\psi$ is identified with $\chi$ in our model. $S$ is the SUSY breaking field, and $m_{3/2}$ denotes the gravitino mass:
\begin{equation}
m_{3/2} \simeq \dfrac{\left|F \right|}{\sqrt{3} M_{pl}}
\end{equation}
with $F$ being the $F$-term VEV of $S$. At this point, $\eta$ is composed of two real scalars, whose mass eigenvalues are as follows: 
\begin{align}
m^2_{\phi_1}&=3 \left[Re\left(\kappa_2\right)+ \left|\kappa_3\right|\right] m_{3/2}^2 \notag \\ &+|ab| \left<H_u^0\right> \left< H_d^0 \right> \cos{(\theta_1-\theta_2)} +|b|^2 v^2_{higgs},
\end{align}
\begin{align}
m^2_{\phi_2}&=3 \left[Re\left(\kappa_2\right)- \left|\kappa_3\right|\right] m_{3/2}^2 \notag \\ &-|ab| \left<H_u^0\right> \left< H_d^0 \right> \cos{(\theta_1-\theta_2)}+|b|^2 v^2_{higgs},
\end{align}
where $\theta_1 =Arg\left(\kappa_3\right), \theta_2=Arg\left(ab\right)$ and $v^2_{higgs}=\left<H_u^0\right>^2+\left<H_d^0\right>^2$.
%It should be noted that while there exists a pseudo Yukawa coupling, the $p$-wave annihilation gives the leading contribution to $\chi \chi \rightarrow \phi_1 \phi_2, \chi \chi \rightarrow \phi_1 \phi_2, \chi \chi \rightarrow W^+ W^-$.
Assuming the following situation:
\begin{equation}Re \left(\kappa_2\right) \simeq -\left|\kappa_3\right| \ \text{or} \ Re\left(\kappa_2\right) \simeq \left|\kappa_3\right|,  
\end{equation}
one obtains the smaller mass:
\begin{align}
\left[\pm |ab| \left<H_u^0\right> \left< H_d^0 \right> \cos{(\theta_1-\theta_2)}+|b|^2 v^2_{higgs}\right]^{1/2},
\end{align}
which lies within $\mathcal{O}(1)$ GeV for $|a|=\mathcal{O}(0.1), |b|=\mathcal{O}(10^{-2})$ and $\tan \beta=20-30$. Besides, postulated $\chi$ is lighter than the gravitino (i.e. $\sqrt{3} \left|\kappa_1 \right| <1$) other than the minimal supersymmetry standard model (MSSM) sparticles, $\chi$ becomes a good DM candidate. Further, it is found that the mixing angle of the lighter boson with $H^0$ is given by \cite{note}
\begin{align}
   \epsilon & = \dfrac{2 \sqrt{3} Re(c) m_{3/2} \cdot v_{higgs} \sin{2 \beta}}{m^2_{H^0}} \notag \\ & \simeq 3.9 \times 10^{-7} \left(\dfrac{Re(c)}{10^{-7}}\right) \left(\dfrac{m_{3/2}}{1\ \text{TeV}}\right) \left(\dfrac{\sin{2 \beta}}{0.1}\right). 
\end{align}

\begin{center}{\bf Conclusion}\end{center}

\noindent
We proposed an approach in which the $p$-wave DM annihilation provides a source of the halo gamma ray signal reported by \cite{Totani:2025fxx}, owing to the Sommerfeld enhancement. Specifically, we considered a weakly coupling of the secluded dark sector with the SM Higgs to show that a pair of the DM annihilation into $W^+W^-$, even though induced at 1-loop level, can be desirably amplified for the DM velocity of $100-200$ km/s. In contrast, at the freeze-out, the interaction within the dark sector yields the correct thermal relics of the DM, while the DM annihilation cross section, for the DM velocity of $\mathcal{O}(10)$ km/s, is compatible with the constraints derived from observations on dwarf galaxies. Besides, it was argued that after the freeze-out, the fermionic DM remains out of thermal equilibrium despite the amplified annihilation into two bosons. Also, we demonstrated that the boson may decay into the SM leptons before Big Bang Nucleosynthesis (BBN). Finally, we suggested that in the language of supersymmetry, the DM may naturally acquire its mass close to $m_{3/2}$, whereas the dark boson can be rendered even lighter, resulting in the appearance of the quasi-bound state.

\begin{center}{\bf Acknowledgment}\end{center}

I would like to thank Dr. S. Biondini (University of Freiburg) for informing me of their work.


\begin{thebibliography}{999}
\bibitem{1}
The scattering rate is roughly given by
\begin{equation}
    \Gamma (=\sigma v n_{eq}) \simeq \dfrac{\lambda^2 T}{16 \pi}\simeq 2.0 \times 10^{-8} \left(\dfrac{\lambda}{10^{-3}}\right)^2 \cdot T.
\end{equation}
at the temperature of $T$. Hence, $\phi$ reaches thermal equilibrium around $\mathcal{O}(10^{10})$ GeV for $|\lambda|\sim 10^{-3}$.
\bibitem{note}
Precisely, $\epsilon$ is somewhat smaller as $\eta$ is formed by a linear combination of
\begin{equation}
    \eta = d_1 \phi_1+d_2 \phi_2
\end{equation}
with $\left|d_1\right|^2+\left|d_2\right|^2=1$.
\end{thebibliography}
\end{document}